\documentclass[sigconf,pbalance]{acmart}
\AtBeginDocument{%
  }

\copyrightyear{2026}
\acmYear{2026}
\setcopyright{cc}
\setcctype{by}
\acmConference[CIKM '26]{Proceedings of the 35th ACM International Conference on Information and Knowledge Management}{November 07--11, 2026}{Rome, Italy}
\acmBooktitle{Proceedings of the 35th ACM International Conference on Information and Knowledge Management (CIKM '26), November 07--11, 2026, Rome, Italy}
\acmDOI{10.1145/3799682.3840118}
\acmISBN{979-8-4007-2539-5/2026/11}

\usepackage{multirow}

\begin{document}

\title{TransRetrieval: Scaling Up Transformer-Based Retrieval for Industrial Recommendation}


\author{Zhifei Zheng}
\orcid{0000-0003-4061-7518}
\authornote{Contributed equally to this research.}
\authornote{Work done during internships at Alibaba Group.}
\email{zhifei.zheng@ruc.edu.cn}
\affiliation{%
  \institution{Gaoling School of Artificial Intelligence, Renmin University of China}
  \city{Beijing}
  \country{China}
}

\author{Yunfei Liu}
\orcid{0000-0001-6377-1513}
\authornotemark[1]
\email{lyf327482@alibaba-inc.com}
\author{Bin Liu}
\orcid{0000-0002-6044-0222}
\authornotemark[1]
\email{zhuoli.lb@alibaba-inc.com}
\affiliation{%
  \institution{Taobao \& Tmall Group of Alibaba}
  \city{Beijing}
  \country{China}
}

\author{Qiren Zhu}
\orcid{0009-0002-2704-0638}
\authornotemark[2]
\email{qiren\_zhu@163.com}
\author{Hanbing Liu}
\orcid{0009-0000-4582-3340}
\authornotemark[2]
\email{liuhanbing@ruc.edu.cn}
\affiliation{%
  \institution{Gaoling School of Artificial Intelligence, Renmin University of China}
  \city{Beijing}
  \country{China}
}

\author{Ziru Xu}
\orcid{0009-0007-4333-1212}
\email{ziru.xzr@alibaba-inc.com}
\author{Han Zhu}
\orcid{0000-0002-9522-5637}
\email{zhuhan.zh@alibaba-inc.com}
\author{Jian Xu}
\orcid{0000-0003-3111-1005}
\email{xiyu.xj@alibaba-inc.com}
\affiliation{%
  \institution{Taobao \& Tmall Group of Alibaba}
  \city{Beijing}
  \country{China}
}

\author{Qi Qi}
\orcid{0000-0001-9192-8928}
\authornote{Corresponding author.}
\additionalaffiliation{%
    \institution{Beijing Key Laboratory of Research on Large Models and Intelligent Governance; Engineering Research Center of Next-Generation Intelligent Search and Recommendation, MOE}
  \city{Beijing}
  \country{China}
}
\email{qi.qi@ruc.edu.cn}
\affiliation{%
  \institution{Gaoling School of Artificial Intelligence, Renmin University of China}
  \city{Beijing}
  \country{China}
}

\author{Bo Zheng}
\orcid{0000-0002-4037-6315}
\authornotemark[3]
\email{bozheng@alibaba-inc.com}
\affiliation{%
  \institution{Taobao \& Tmall Group of Alibaba}
  \city{Beijing}
  \country{China}
}

\renewcommand{\shortauthors}{Zheng, Liu, Liu et al.}


\begin{abstract}
Applying scaling laws to recommendation retrieval is hindered by feature heterogeneity: naively stacking Transformer layers yields diminishing returns because heterogeneous fields produce severe token-norm divergence.
We present \textbf{TransRetrieval}, a Transformer-based retrieval framework that scales with both computational budget and cross-domain data.
The key enabler is (1)~weighted average aggregation, which restores the homogeneous-token assumption Transformers rely on.
Building on this, we introduce (2)~target token compression that cuts per-candidate FLOPs by 85\% while preserving cross-attention expressiveness, and (3)~position-style domain embeddings that unify multiple domains at negligible additional cost, turning cross-domain data into a scaling asset.
On a 40-billion-interaction industrial dataset and the public KuaiRand benchmark, scaling compute from 0.1 to 2\,MFLOPs per target yields +19.3\,/\,+22.2\,pt Recall@2000, confirming robust log-linear scaling.
In online A/B tests, TransRetrieval lifts platform revenue by 2.53\% under the same end-to-end latency constraint as the production baseline.
\end{abstract}

\ccsdesc[500]{Information systems~Search engine architectures and scalability}
\ccsdesc[500]{Information systems~Recommender systems}
\ccsdesc[300]{Computing methodologies~Learning paradigms}

\keywords{Large-scale Retrieval, Transformer, Scaling Law, Embedding Normalization, Multi-domain Modeling, Recommendation System}



\maketitle

\section{Introduction}
\label{sec:introduction}

Scaling laws have been the driving force behind the rapid progress of large language models, where performance improves predictably with model size, data, and compute~\cite{DBLP:journals/corr/abs-2203-15556}.
Recommendation systems, despite their massive datasets and growing computational budgets, have largely failed to exhibit analogous scaling behavior.
Recent studies confirm this gap: traditional deep recommendation models yield diminishing or even negative returns when simply scaled up~\cite{DBLP:conf/icml/ZhangLCNLLZHYWP24,DBLP:conf/cikm/XuWGGXHWL25}.
Transformers, whose power-law scaling properties are well established in NLP and vision, are natural candidates for closing this gap.

Nevertheless, scaling Transformers in large-scale retrieval, where candidate pools reach up to $10^{8}$ items and latency budgets are measured in tens of milliseconds, faces several challenges:
\textbf{(1) Heterogeneous feature scale misalignment.}
Standard Transformers process homogeneous input tokens: in NLP every word embedding is drawn from the same lookup table~\cite{DBLP:conf/nips/VaswaniSPUJGKP17}; in vision every patch passes through an identical projection~\cite{DBLP:journals/corr/abs-2010-11929}.
Recommendation features, however, are inherently heterogeneous (user profiles, behavioral sequences, and item attributes), and mapping them into Transformer tokens produces severe scale misalignment~\cite{DBLP:conf/cikm/ZhuFZJWHDWZGYCC25}.
This misalignment silently degrades representation quality by structurally distorting the embedding space.
\textbf{(2) Tight computation budget versus latency constraints.}
Retrieval must serve corpora of up to $10^{8}$ items under strict queries per second (QPS) expectation ~\cite{DBLP:conf/recsys/CovingtonAS16,DBLP:journals/corr/abs-2407-21022}, where the constraint is orders of magnitude more stringent than in ranking~\cite{DBLP:conf/cikm/BorisyukSZPAPBP24}.
\textbf{(3) Scaling cost multiplication across domains.}
Industrial systems routinely serve multiple business domains (e.g., homepage, search, detail page) with distinct behavioral distributions~\cite{DBLP:conf/cikm/JiangLZYLXDZ22}.
Maintaining a separate deep model per domain multiplies both training and serving cost, and sparse domains lack the data to benefit from scaling alone. Naively merging domains, however, fails because their distributions conflict.

Several recent efforts advance scaling in recommendation, including HSTU~\cite{DBLP:conf/icml/ZhaiLLWLCGGGHLS24}, RankMixer~\cite{DBLP:conf/cikm/ZhuFZJWHDWZGYCC25}, MARM~\cite{DBLP:conf/cikm/LvCGZQZ0Z25}, Climber~\cite{DBLP:conf/cikm/XuWGGXHWL25} and MTGR~\cite{DBLP:conf/cikm/HanYCJJ0MHLJHZY25}. Each of them either flattens or pre-groups the heterogeneous feature set, and all of them target ranking, which scores only the candidates retrieval has already shortlisted.
How to scale Transformers for retrieval over corpora of up to $10^{8}$ items while preserving heterogeneous features under strict latency budgets thus remains largely unexplored.

We present \textbf{TransRetrieval}, a Transformer-based retrieval framework deployed in the Alibaba display advertising platform, serving users across 4 business domains.
Three design choices, each addressing one of the challenges above, jointly enable scaling:
\textbf{(1) Weighted average aggregation} resolves scale misalignment at the source by decoupling embedding norms from feature cardinality, restoring the homogeneous-token assumption that Transformers rely on.
\textbf{(2) Target token compression} projects all target-side features into a single $D$-dimensional token via a lightweight multilayer perceptron (MLP), reducing per-candidate FLOPs by 85\% and enabling Hierarchical Navigable Small World (HNSW) based sub-linear retrieval.
Inference cost is cut further by sharing one KV cache across targets: since all candidates in a request are scored against the same user, the user-side keys and values are computed once per request and reused by every candidate.
\textbf{(3) Position-style domain embedding} injects domain information via element-wise addition, analogous to positional encoding. It unifies 4 domains into a single model at negligible additional FLOPs and turns multi-domain data into a scaling asset, with sparse domains benefiting from cross-domain transfer.

All three designs use only standard operators and no extra infrastructure, offering a transferable recipe for deploying Transformers in retrieval.
Deployed in the Alibaba display advertising platform, TransRetrieval achieves a 2.53\% platform revenue lift in online A/B tests; scaling compute from 0.1 to 2\,MFLOPs per target yields +19.3\,pt Recall@2000 on the industrial dataset and +22.2\,pt on the public KuaiRand benchmark, confirming robust log-linear scaling.
To our knowledge, this is among the first systematic validations of scaling laws in the retrieval stage of the recommendation architecture, complementing recent ranking-stage results demonstrated by Climber~\cite{DBLP:conf/cikm/XuWGGXHWL25}.

\section{Related Work}

\textbf{Scaling in recommendation and the heterogeneous feature barrier.}
Wukong~\cite{DBLP:conf/icml/ZhangLCNLLZHYWP24} provides the first systematic diagnosis of scaling inefficiency in deep recommendation models, attributing it to upscaling mechanisms that expand embedding tables without enhancing feature interaction.
Subsequent efforts bypass the heterogeneity barrier rather than resolve it:
HSTU~\cite{DBLP:conf/icml/ZhaiLLWLCGGGHLS24} flattens all features into a single behavioral sequence, sacrificing structural distinction between feature types;
RankMixer~\cite{DBLP:conf/cikm/ZhuFZJWHDWZGYCC25} pre-groups features into fixed semantic channels processed by parameter-free token mixing, while MARM~\cite{DBLP:conf/cikm/LvCGZQZ0Z25} reaches comparable gains only with 60\,TB of cache infrastructure;
Climber~\cite{DBLP:conf/cikm/XuWGGXHWL25} focuses on multi-scale sequence extraction where inputs are behavioral tokens from the same embedding table, and MTGR~\cite{DBLP:conf/cikm/HanYCJJ0MHLJHZY25} retains rich features under the generative paradigm.
All of these validate scaling at the ranking stage only; how to enable stable deep Transformer scaling in retrieval while retaining the full heterogeneous feature set remains unresolved.

\textbf{Model-based retrieval and Transformer efficiency.}
NANN~\cite{DBLP:conf/cikm/ChenLZWLMHJXDZ22} establishes the model-based retrieval paradigm: a deep neural network (DNN) scoring function drives beam search over an HNSW graph~\cite{DBLP:journals/pami/MalkovY20}, so that only a small fraction of the corpus is ever scored: the model itself determines which candidates are evaluated.
BAR~\cite{BAR}, which serves as our production baseline, extends this paradigm with bidding-awareness for multi-stage consistency, but keeps the scoring network small.
On the user side, the evolution from two-tower architectures~\cite{DBLP:conf/recsys/CovingtonAS16} to sequential Transformers (SASRec~\cite{DBLP:conf/icdm/KangM18}, BERT4Rec~\cite{DBLP:conf/cikm/SunLWPLOJ19}) has enriched user representations, and KuaiFormer~\cite{DBLP:journals/corr/abs-2411-10057} reformulates retrieval as next-action prediction under a generative paradigm.
Under the model-based paradigm, however, each candidate still requires a forward pass through the target network, making per-candidate FLOPs the dominant bottleneck, which is why the scoring network has stayed small.
TransRetrieval stays within this paradigm but relieves that bottleneck: compressing each candidate to a single token makes the per-candidate pass cheap enough to scale the scoring network itself.

\textbf{Multi-domain recommendation and cross-domain data scaling.}
Expert-based methods (MMoE~\cite{DBLP:conf/kdd/MaZYCHC18}, PLE~\cite{DBLP:conf/recsys/TangLZG20}) share expert parameters across domains, but gate-based soft routing means sparse domains, whose gates are undertrained due to limited data, cannot fully leverage the shared capacity for effective data scaling.
STAR~\cite{DBLP:conf/cikm/ShengZZDDLYLZDZ21} unifies serving through shared and domain-specific components, but its star topology has not been shown to enable cross-domain transfer that benefits data-scarce domains.

\section{TransRetrieval: Methodology}
\label{sec:methodology}

\begin{figure*}[t]
    \centering
    \includegraphics[width=1\linewidth]{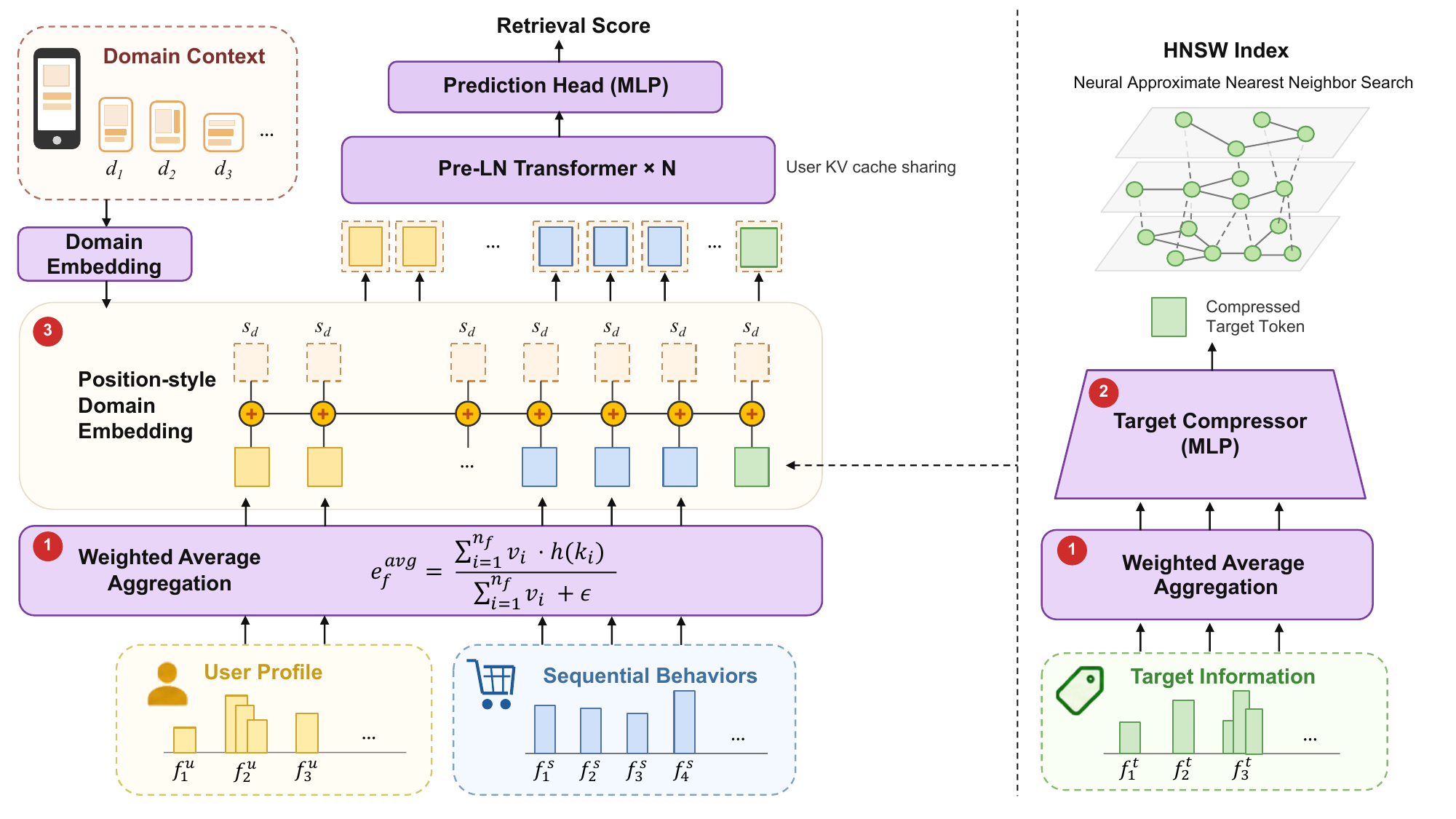}
    \caption{The overall architecture of TransRetrieval.
     All three pillars are lightweight designs: (i) Heterogeneous features are first aggregated via weighted average. (ii) Target features are compressed into a single token by an MLP. (iii) Domain context is injected via position-style embeddings (added to all tokens). A shared Transformer then scores every candidate of a request against one user-side KV cache, computed once and reused across all targets.
    }
    \Description{An architecture diagram showing the data flow of TransRetrieval. Heterogeneous user-side and target-side feature fields are first aggregated via weighted average into per-field tokens. Target-side tokens are then compressed into a single token by a 3-layer MLP, while user-side tokens are kept one-per-field. A learnable domain embedding is added element-wise to every token. The resulting tokens enter a shared Transformer; user-side KV pairs are cached once and reused across all target candidates during scoring.}
    \label{fig:TR-overall}
\end{figure*}

We first formalize the retrieval paradigm and overview the architecture in Section~\ref{subsec:preliminaries}, then detail our three core designs in Sections~\ref{subsec:weighted_avg}--\ref{subsec:domain_embedding}.

\subsection{Preliminaries}
\label{subsec:preliminaries}

\textbf{Model-based Retrieval.}
TransRetrieval operates under the model-based retrieval paradigm established by NANN~\cite{DBLP:conf/cikm/ChenLZWLMHJXDZ22}: a DNN scoring function $s(u,i) = F_\theta(u,i)$ drives beam search over an HNSW graph~\cite{DBLP:journals/pami/MalkovY20}, which visits only a small fraction $\rho$ of the corpus per request.
We express the scoring budget as this fraction rather than as an absolute count, because it is $\rho$, not the corpus size, that determines the per-request cost: for a corpus of $N$ items, the model performs $\rho N$ forward passes instead of $N$.
Practical settings put $\rho$ at roughly $0.1\%$--$1\%$; our deployment scores ${\sim}4\times10^{4}$ candidates over a corpus of $52$\,M items ($\rho \approx 0.08\%$), and the same pipeline serves corpora up to the $10^{8}$ scale.
Unlike two-tower models where $s(u,i) = \langle f(u), g(i) \rangle$ restricts interaction to inner product, model-based retrieval permits arbitrary cross-feature interaction at the cost of a per-candidate forward pass, which makes per-candidate FLOPs the dominant bottleneck.

\textbf{Feature Representation.}
Each feature field $f$ is represented as a set of (key, weight) pairs $\{(k_i, v_i)\}_{i=1}^{n_f}$, where $k_i$ indexes to an embedding table $h: \mathcal{K} \to \mathbb{R}^D$ and $v_i \in \mathbb{R}$ is its weight.

\textbf{Architecture Overview.}
As illustrated in Figure~\ref{fig:TR-overall}, TransRetrieval processes features through four stages:
(1)~each feature field is aggregated via weighted average (Sec.~\ref{subsec:weighted_avg}), producing Transformer-friendly tokens;
(2)~target-side tokens are compressed into a single $D$-dimensional token via a lightweight MLP (Sec.~\ref{subsec:target_compression}), while each user-side feature forms $1$ token;
(3)~a learnable domain embedding is added element-wise to all tokens (Sec.~\ref{subsec:domain_embedding});
(4)~the resulting tokens enter an $N_\text{layer}$-layer Pre-LN Transformer with ReLU feed-forward network (FFN), where the user-side key–value (KV) pairs of a request are cached once and reused across all of its target candidates.

\subsection{Bridging Heterogeneous Features via Weighted Average Aggregation}
\label{subsec:weighted_avg}

Profiling our model under conventional \textit{weighted-sum} aggregation reveals severe scale misalignment: user-side token norms reach up to $10\times$ the target-side norms.
This imbalance causes attention weights to be dominated by high-norm tokens.
The root cause is that \textit{weighted sum} aggregation produces output norms growing with feature cardinality $n_f$: user-side fields with multi-value features inherently accumulate much larger norms than target-side fields.

We resolve this with \textbf{weighted average aggregation}:
\begin{equation}
    \mathbf{e}_f = \frac{\sum_{i=1}^{n_f} v_i \cdot h(k_i)}{\sum_{i=1}^{n_f} v_i + \epsilon}
    \label{eq:weighted_avg}
\end{equation}
Division by $\sum_i v_i$ simultaneously cancels both the $n_f$-dependent accumulation and the heterogeneous weight magnitudes, bounding the output norm solely by the embedding table's distribution regardless of field cardinality.

\textbf{Why not alternative aggregators?} \textit{Max pooling} introduces an extreme-value bias that grows as $O(\sqrt{\ln n_f})$ and discards the majority of feature information.
\textit{LayerNorm}~\cite{DBLP:journals/corr/BaKH16} applied post-aggregation underperforms empirically (Sec.~\ref{sec:wavg_exp}): its centering and rescaling erase the per-field magnitude signal that distinguishes heterogeneous fields.
\textit{Weighted average} is an operator that simultaneously (a)~bounds output scale independently of $n_f$ and (b)~requires no learned or estimated statistics: it is a closed-form linear combination that remains stable regardless of $D$.

\subsection{Efficient Retrieval via Target Token Compression}
\label{subsec:target_compression}

Because all candidates of a request share one user, the user-side K/V are computed once per request and reused by every target; each target candidate then only requires its own Q/K/V/output projections plus FFN ($16L_t D^2$ per layer) plus cross-attending to the full user sequence of length $L_u$ ($4L_t L_u D$ per layer). Summing across $N_\text{layer}$ layers and $N_\text{target}$ candidates:
\begin{equation}
    \text{FLOPs}_\text{total} \approx 4D \cdot N_\text{target} \cdot L_t \cdot (4D + L_u) \cdot N_\text{layer}
    \label{eq:flops}
\end{equation}
where $L_t$ is the number of target tokens and $L_u$ the user sequence length.
Among these factors, $D$ and $N_\text{layer}$ are scaling parameters we wish to maximize; $N_\text{target} = \rho N$ is already reduced by HNSW to a small fraction of the corpus; and $L_u$ is dominated by the $4D$ term.
This leaves $L_t$ as the only factor with meaningful room for reduction. In the uncompressed model, $L_t$ scales with the number of target feature fields, making it a natural target for compression.

We introduce a \textbf{target compressor}: a 3-layer MLP with hidden sizes $[8D, 4D, D]$, Parametric ReLU activations and LayerNorm between layers (final layer is a plain linear projection), that maps the concatenation of all target-side aggregated embeddings into a single token:
\begin{equation}
    \mathbf{t} = \text{MLP}\!\left(\left[\mathbf{e}_{f_1}, \mathbf{e}_{f_2}, \ldots, \mathbf{e}_{f_M}\right]\right) \in \mathbb{R}^D
    \label{eq:target_compression}
\end{equation}
Crucially, this compression is not a compromise but a strategic reallocation: the freed FLOPs are reinvested into scaling $D$ and $N_\text{layer}$, enabling deeper and wider architectures within the same latency envelope, trading redundant per-candidate tokens for richer cross-feature interactions.

The single compressed token also naturally serves as a fixed-dimensional vector for HNSW indexing without further processing.

\subsection{Cross-Domain Data Scaling via Position-Style Domain Embedding}
\label{subsec:domain_embedding}

As motivated in Sec.~\ref{sec:introduction}, a unified multi-domain model is required to avoid per-domain cost multiplication. The remaining design question is \emph{how} to inject domain identity. A straightforward strategy is appending $N_t$ extra tokens encoding the domain to the input sequence, but this inflates attention cost from $O(L^2)$ to $O((L{+}N_t)^2)$.

We instead introduce a \textbf{position-style domain embedding}: a learned vector $\mathbf{s}_d \in \mathbb{R}^D$ indexed by domain $d$, added element-wise to every token:
\begin{equation}
    \tilde{\mathbf{e}}_f = \mathbf{e}_f + \mathbf{s}_d
    \label{eq:domain_embedding}
\end{equation}
Analogous to positional encoding~\cite{DBLP:conf/nips/VaswaniSPUJGKP17}, domain identity is injected without increasing sequence length or attention complexity.

The key insight is that this design enables \textbf{cross-domain data scaling}: because domain information enters only through element-wise addition, all Transformer parameters, including every attention head and FFN layer, are unconditionally shared across domains.
Every training sample from every domain contributes gradients to the same parameters, so sparse domains are no longer data-starved.
This contrasts with MMoE/PLE where gate routing implicitly partitions gradient flow.

\subsection{System Implementation}
\label{subsec:system}

Deploying TransRetrieval in production requires co-designing the inference system with the model architecture to meet strict latency constraints.

\textbf{Hardware-aware inference optimization.}
A request's user-side KV pairs are computed once and shared across the scoring passes of all its candidates. Standard LLM inference frameworks would replicate that cache once per candidate sequence; we instead replace the cache manager with an expand operator that broadcasts the single user cache to every candidate without copying, reducing per-pass latency by 30\%.
After target compression, each candidate is a single query token attending to the full user context. This extreme aspect ratio severely underutilizes modern GPU hardware, which is optimized for large matrix operations rather than single-row queries.
We pack multiple candidates into a single batched computation with masking to preserve per-candidate isolation, significantly improving hardware utilization.
Combined with operator fusion across LayerNorm, residual, and activation layers, the model forward latency drops by $11\times$ vs.\ naive implementation.

\textbf{GPU retrieval.}
Model-based retrieval interleaves scoring with graph search across beam-search iterations.
When the index resides on CPU, each iteration incurs multiple round-trip host-device transfers that fragment the GPU computation graph.
We migrate all retrieval operators (neighbor lookup, distance computation, and candidate selection) onto GPU, unifying scoring and search into a single device-resident pipeline and reducing retrieval latency by 89\% vs.\ CPU retrieval.
However, co-locating the full index on GPU introduces memory pressure.
We reduce GPU memory through inference engine sharing across requests and quantizing static index data, increasing serving concurrency by $3\times$.

\textbf{Near-line update.}
The target compressor (Sec.~\ref{subsec:target_compression}) is architecturally decoupled from the Transformer and deployed as an independent sub-graph.
During full deployment, this sub-graph computes representations for all targets to build the retrieval index.
To maintain freshness, an event-driven near-line service monitors target feature mutations and invokes the same sub-graph to recompute only the affected representations, propagating updates to the live index asynchronously.

\section{Experiments}
\label{sec:experiments}

We focus on the following research questions:
\begin{itemize}
    \item \textbf{RQ1}: How does TransRetrieval compare to state-of-the-art baselines under similar computational budgets? (Sec.~\ref{sec:overall})
    \item \textbf{RQ2}: Does TransRetrieval exhibit predictable scaling behavior as model capacity increases? (Sec.~\ref{sec:scaling})
    \item \textbf{RQ3}: How does each of the three design pillars contribute to performance? (Sec.~\ref{sec:ablation} )
    \item \textbf{RQ4}: Does TransRetrieval deliver significant improvements in real-world deployment? (Sec.~\ref{sec:online})
\end{itemize}

\subsection{Experimental Protocol}

\subsubsection{Datasets}

\begin{table*}[!t]
\caption{Statistics of the two datasets. \#f\_user and \#f\_target denote the number of user-side and target-side features; Len\_seq denotes the length of user behavior sequences.}
\centering
\small
\setlength{\tabcolsep}{5pt}
\begin{tabular}{lcccccccrrrr}
\toprule
\multirow{2}{*}{Dataset} & 
\multirow{2}{*}{\#Domains} & 
\multirow{2}{*}{\#Users} & 
\multirow{2}{*}{\#Targets} & 
\multirow{2}{*}{\#Interactions} & 
\multirow{2}{*}{\#f\_user} & 
\multirow{2}{*}{\#f\_target} & 
\multirow{2}{*}{Len\_seq} &
\multicolumn{4}{c}{Domain Distribution (\%)} \\
\cline{9-12}
& & & & & & & & Dom A/0 & Dom B/1 & Dom C/2 & Dom D/3 \\
\midrule
Industrial     & 4 & 250M & 52M & 40B & 75 & 87 & 100 & 32\% & 20\% & 26\% & 22\% \\
KuaiRand       & 4 & 24,943 & 536,491 & 152M & 37 & 63 & 100 & 11.7\% & 84.8\% & 3.32\% & 0.13\% \\
\bottomrule
\end{tabular} 
\label{tab:dataset_desc}
\end{table*}

We conduct experiments on two multi-domain datasets from two different recommender system cases: display advertisement and short video recommendation.

\begin{itemize}
    \item \textbf{Industrial}. A large-scale industrial dataset sampled from the production environment of the Alibaba display advertising platform, containing $40$ billion interactions over a corpus of $52$\,M advertised items. The dataset covers 4 distinct business domains (anonymized as domains A, B, C, and D), capturing diverse user–item interaction patterns. It includes rich user behavioral histories, detailed item attributes, and real-time contextual signals.
    
    \item \textbf{KuaiRand}~\cite{DBLP:conf/cikm/GaoLZCLL0022}. A publicly available sequential recommendation dataset from Kuaishou App, featuring 12 types of user feedback signals and comprehensive interaction histories. Following previous retrieval works~\cite{DBLP:conf/www/ZhangDDLJG23,DBLP:conf/www/WangXHSZW25}, we filter the dataset with a frequency threshold of 1000 click records for users and 100 exposure records for items. We use the first 4 domains (tab=0, 1, 2, 3) in our experiments.
\end{itemize}

Table~\ref{tab:dataset_desc} presents dataset statistics. 
Both datasets cover four domains with diverse user behaviors. In Industrial, Domain C aggregates multiple smaller sub-domains with heterogeneous user behavior patterns, resulting in significant internal distributional variance. KuaiRand includes an extremely sparse domain (Tab 3), which provides a challenging testbed for generalization under data scarcity.
Additionally, the two datasets differ in numbers of attributes, which contributes to the different computational costs observed in subsequent experiments.
For both datasets, we adopt a temporal split strategy to ensure no data leakage occurs, simulating realistic deployment domains where models are trained on historical data and evaluated on future interactions.

\subsubsection{Evaluation Metrics}

We focus on Recall@2000 (abbreviated as R@2000 when space is limited), the proportion of ground-truth positive items appearing in the top-2000 retrieved candidates.  

\subsubsection{Baselines}

We include the following baselines:
\begin{itemize}
    \item \textbf{Production Baseline}~\cite{BAR}. The bidding-aware model-based retrieval system serving the Alibaba display advertising platform before this work: a highly optimized industrial standard, and the system TransRetrieval replaces in our online A/B test.

    \item \textbf{KuaiFormer}~\cite{DBLP:journals/corr/abs-2411-10057}. A simple two-tower retrieval baseline that encodes the user side with a Transformer and scores candidates via inner-product search.

    \item \textbf{HSTU}~\cite{DBLP:conf/icml/ZhaiLLWLCGGGHLS24}. Flattens all features into a single behavioral sequence to achieve token homogeneity, discarding structural feature distinctions.

    \item \textbf{RankMixer}~\cite{DBLP:conf/cikm/ZhuFZJWHDWZGYCC25}. A scalable ranking architecture adapted for retrieval; replaces attention with parameter-free token mixing and concentrates computation in per-token FFNs.
\end{itemize}

For fair comparison, we implement all baselines within our framework, sharing the same preprocessing pipeline. Each baseline retains its original feature processing design. 

\subsubsection{Implementation Details}
All experiments are conducted on NVIDIA H20 GPU clusters using PyTorch and RecIS~\cite{recis}. All models and both datasets share one tuned hyperparameter set: AdamW, learning rate $1\text{e-}3$, weight decay $3\text{e-}5$, and the pairwise learning-to-rank (LTR) loss of the production baseline~\cite{BAR,COPR}. Negative sampling ratios are dataset-specific: $5$ negatives per positive for Industrial and $200$ for KuaiRand, reflecting training data scale: the much larger Industrial volume (40\,B vs.\ 152\,M) yields sufficient negative signal at a low ratio, while KuaiRand requires a higher ratio to compensate for limited positives. All runs train to convergence, and the final checkpoint is evaluated.

\begin{table*}[t]
    \caption{
    Overall retrieval performance (Recall@2000) across datasets and domains. \textbf{Overall} is the per-domain Recall weighted by each domain's share of evaluation interactions. Among baselines and TransRetrieval-128D5L, best results are in \textbf{bold}, second-best underlined.
    }
    \setlength{\tabcolsep}{3pt} 
    \centering
    \small
    \begin{tabular}{lcccccl|cccccl}
    \toprule
    \multirow{2}{*}{\textbf{Method}}
    & \multicolumn{6}{c|}{\textbf{Industrial (Recall@2000)}}
    & \multicolumn{6}{c}{\textbf{KuaiRand (Recall@2000)}} \\
    \cmidrule(lr){2-7} \cmidrule(lr){8-13}
    & Domain A & Domain B & Domain C & Domain D & Overall & MFLOPs$^{'}$
    & Tab 0 & Tab 1 & Tab 2 & Tab 3 & Overall & MFLOPs$^{'}$ \\
    \midrule
    KuaiFormer-128D5L\footnotemark
      & 0.496  & 0.466  & 0.470  & 0.548  & 0.495  & 0.011
      & 0.371  & \underline{0.415}  & 0.379  & 0.463  & \underline{0.409}  &  0.012 \\
    \midrule
    Production Baseline & \underline{0.572} & \underline{0.586} & \underline{0.548} & 0.608 & \underline{0.576} & 0.69 & 0.401 & 0.378 & 0.299 & 0.500 & 0.378 & 0.64 \\
    \midrule
    HSTU-128D5L & 0.518 & 0.490 & 0.474 & \underline{0.636} & 0.527 & 1.72 & 0.315 & 0.394 & 0.312 & \textbf{0.611} & 0.382 & 1.72 \\
    RankMixer-128D8T & 0.445 & 0.419 & 0.437 & 0.512 & 0.452 & 2.10 & \underline{0.406} & 0.357 & \underline{0.384} & 0.518 & 0.364 & 2.10 \\
    \cmidrule(lr){1-13}
    TransRetrieval-64D3L & 0.593 & 0.590 & 0.575 & 0.663 & 0.603 & 0.45 & 0.460 & 0.413 & 0.413 & 0.481 & 0.419 & 0.42 \\
    TransRetrieval-128D5L & \textbf{0.644} & \textbf{0.641} & \textbf{0.634} & \textbf{0.719} & \textbf{0.657} & 1.91 & \textbf{0.526} & \textbf{0.465} & \textbf{0.479} & \underline{0.537} & \textbf{0.473} & 1.81 \\
    \bottomrule
    \end{tabular}
    \label{tab:reconstructed_performance_mflops}

\end{table*}



\subsection{Overall Performance (RQ1)}
\label{sec:overall}

We compare the overall retrieval performance under similar backbone configurations across both datasets and all domains. For model configurations, we use the notation $xD\text{-}yL$ where $x$ denotes the embedding dimension and $y$ the number of Transformer layers. Table~\ref{tab:reconstructed_performance_mflops} presents results comparing TransRetrieval (128D-5L, 64D-3L) against the production baseline, HSTU (128D-5L), KuaiFormer (128D-5L), and RankMixer (128D-8T, where T denotes the number of token mixing blocks). MFLOPs$^{'}$ measures the per-candidate compute on the target side (Millions of FLOPs per target).

Under a comparable per-candidate budget, TransRetrieval-64D3L (0.45 MFLOPs$^{'}$) already surpasses the highly optimized Production Baseline (0.69 MFLOPs$^{'}$) across all domains, achieving 0.603 vs.\ 0.576 Overall Recall@2000 with 35\% fewer FLOPs.
Scaling up to TransRetrieval-128D5L (1.91 MFLOPs$^{'}$) further lifts performance to 0.657, a gain of +5.4\,pt over the smaller configuration, demonstrating that additional compute translates reliably into retrieval quality.
Against the scalable baselines HSTU and RankMixer at similar per-candidate budgets (\textasciitilde{}1.7--2.1 MFLOPs$^{'}$), TransRetrieval-128D5L sets a new state-of-the-art, and even TransRetrieval-64D3L surpasses both at approximately one quarter of their cost.
KuaiFormer trades expressiveness for amortized cost; its scaling behavior is examined in Sec.~\ref{sec:scaling}.

\subsection{Scaling Behavior (RQ2)}
\label{sec:scaling}

\begin{figure}[t]
    \centering
    \includegraphics[width=\linewidth]{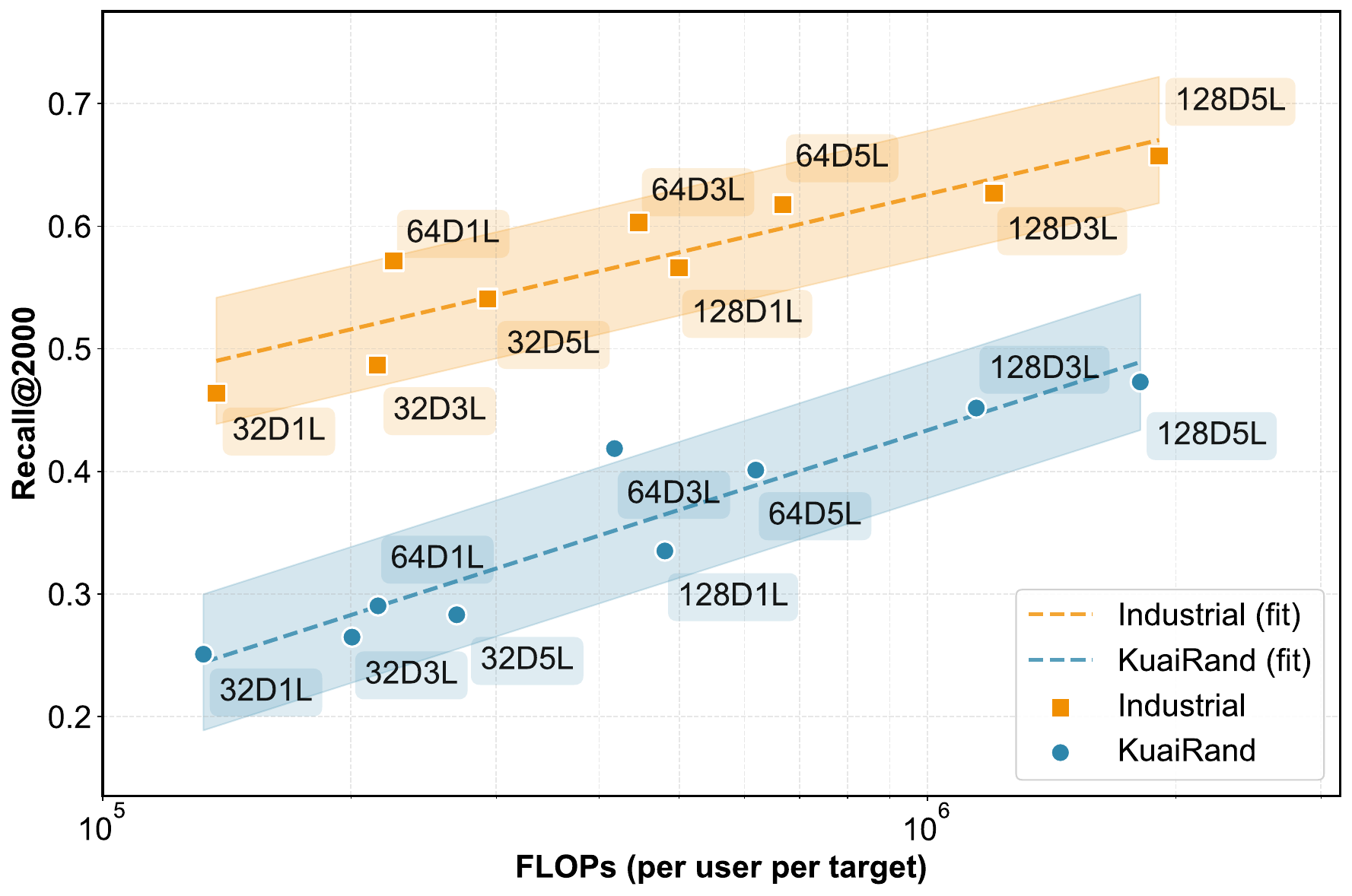}
    \caption{Scaling curves of TransRetrieval. Each point is one $(D, N_\text{layer})$ configuration; shaded regions are 95\% confidence intervals of the log-linear fits.}
    \Description{A scatter plot with two log-linear scaling curves overlaid on the same axes. The horizontal axis is per-target FLOPs on a logarithmic scale; the vertical axis is Recall@2000. Each point corresponds to a (dimension, depth) configuration. The Industrial curve rises monotonically from about 0.46 at 0.14 MFLOPs to 0.66 at 1.91 MFLOPs; the KuaiRand curve rises from about 0.25 to 0.47 over the same FLOPs range. Both curves are tightly fit by log-linear functions ($R^2 \approx 0.82$ and $0.88$), with shaded 95\% confidence bands.}
    \label{fig:scaling_curves}
\end{figure}

TransRetrieval unlocks predictable scaling for retrieval models.
We sweep embedding dimensions (32D, 64D, 128D) and layer counts (1L, 3L, 5L), and plot Recall@2000 against per-target FLOPs in Figure~\ref{fig:scaling_curves}.
Both datasets exhibit clean log-linear scaling: Recall@2000 improves from 0.464 to 0.657 on Industrial (+19.3\,pt) and from 0.251 to 0.473 on KuaiRand (+22.2\,pt) as the backbone grows from 32D-1L to 128D-5L, with high $R^2$ ($0.82$/$0.88$) confirming the trend is consistent and predictable.

Table~\ref{tab:scaling_all} reports detailed results across model sizes.
Most strikingly, our 128D-5L (1.91\,MFLOPs$^{'}$) outperforms RankMixer-768D-16T by 12.6\,pt on Industrial while using nearly two orders of magnitude less compute (1.91 vs.\ 150.99\,MFLOPs$^{'}$).

KuaiFormer shows near-flat scaling (0.485$\to$0.495 when doubling dimension), confirming that the two-tower inner-product bottleneck cannot be overcome by enlarging the encoder alone.

\begin{table}[t]
    \caption{Scaling performance of all methods across different configurations on both datasets (R@2000). MFLOPs$^{'}$ denotes per-candidate scoring cost for each dataset.
    }
    \centering
    \small
    \setlength{\tabcolsep}{2pt} 
    \begin{tabular}{crrrrr}
    \toprule
    & & \multicolumn{2}{c}{\textbf{Industrial}} & \multicolumn{2}{c}{\textbf{KuaiRand}} \\
    \cmidrule(lr){3-4} \cmidrule(lr){5-6}
    \textbf{Model} & \textbf{Config} & \textbf{R@2000} & \textbf{MFLOPs$^{'}$} & \textbf{R@2000} & \textbf{MFLOPs$^{'}$} \\
    \midrule
    \multirow{2}{*}{KuaiFormer\footnotemark[\value{footnote}]} & 64D5L & 0.485 & 0.003 & 0.357 & 0.004 \\
    & 128D5L & 0.495 & 0.011 & 0.409 & 0.012 \\
    \midrule
    \multirow{4}{*}{HSTU} & 32D1L & 0.380 & 0.13 & 0.269 & 0.13 \\
    & 64D2L & 0.437 & 0.30 & 0.298 & 0.30 \\
    & 128D3L & 0.509 & 1.09 & 0.328 & 1.09 \\
    & 128D5L & 0.527 & 1.72 & 0.382 & 1.72 \\
    \midrule
    \multirow{3}{*}{RankMixer} & 128D8T & 0.452 & 2.10 & 0.364 & 2.10 \\
    & 768D8T & 0.468 & 75.50 & 0.420 & 75.50 \\
    & 768D16T & 0.531 & 150.99 & 0.453 & 150.99 \\
    \midrule
    \multirow{3}{*}{TransRetrieval} & 32D1L & 0.464 & 0.14 & 0.251 & 0.13 \\
    & 64D3L & 0.603 & 0.45 & 0.419 & 0.42 \\
    & 128D5L & 0.657 & 1.91 & 0.473 & 1.81 \\
    \bottomrule
    \end{tabular}
    \label{tab:scaling_all}

\end{table}
\subsection{Ablation Study (RQ3)}
\label{sec:ablation}

\footnotetext{KuaiFormer is a two-tower model, here we report its MFLOPs$^{'}$ as the per-user encoding cost amortized over the ${\sim}$40k (Industrial) / ${\sim}$28k (KuaiRand) candidates actually scored during HNSW graph traversal.}

\subsubsection{Effect of Weighted Average Aggregation}
\label{sec:wavg_exp}
\begin{table}[t]
    \caption{Recall@2000 of different feature aggregation strategies. Weighted Average achieves the best performance.}
    \label{tab:wavg_ablation}
    \centering
    \small
    \setlength{\tabcolsep}{2pt}
    \begin{tabular}{lccccc}
    \toprule
    \textbf{Method} & \textbf{Dom A} & \textbf{Dom B} & \textbf{Dom C} & \textbf{Dom D} & \textbf{Overall} \\
    \midrule
    Weighted Average (ours) & \textbf{0.610} & \textbf{0.598} & \textbf{0.589} & \textbf{0.681} & \textbf{0.618} \\
    Weighted Sum            & 0.600 & 0.594 & 0.572 & 0.674 & 0.608 \\
    LN + Weighted Sum       & 0.582 & 0.575 & 0.568 & 0.656 & 0.593 \\
    Max Pooling             & 0.563 & 0.554 & 0.533 & 0.644 & 0.571 \\
    \bottomrule
    \end{tabular}
\end{table}

We compare our weighted average aggregation against alternative feature fusion strategies in Table~\ref{tab:wavg_ablation}, all using the same 64D-5L configuration.
As shown in Table~\ref{tab:wavg_ablation}, the weighted average consistently outperforms all alternatives across every domain.
Plain Weighted Sum lags behind (0.608 vs.\ 0.618) due to unbounded norm growth as the number of features increases.
Adding LayerNorm after weighted summation (LN + Weighted Sum) also hurts substantially (0.593 vs.\ 0.618), because its centering and rescaling erase the per-field magnitude signal that the Transformer relies on to distinguish heterogeneous fields.
Max Pooling suffers the largest degradation (0.571 vs.\ 0.618), because its extreme-value selection discards the majority of feature information.

\subsubsection{Effect of Target Token Compression}
\label{sec:compression}

\begin{table}[t]
    \caption{Impact of target token compression on retrieval performance and computational cost.}
    \centering
    \setlength{\tabcolsep}{3pt}
    \centering
    \small
    \begin{tabular}{lccc}
    \toprule
    \textbf{Config} & \textbf{\# Tokens} & \textbf{R@2000} & \textbf{MFLOPs$^{'}$} \\
    \midrule
    64D-3L  & 8 tokens & 0.624 & 3.04 \\
    64D-3L  & 4 tokens & 0.615 & 1.55 \\
    64D-3L  & 1 token & 0.603 & 0.45 \\
    \midrule
    128D-5L  & 1 token & \textbf{0.657} & 1.91 \\
    \bottomrule
    \end{tabular}
    \label{tab:target_compression}
\end{table}

Table~\ref{tab:target_compression} presents results under the 64D-3L and 128D-5L configurations.
Within 64D-3L, compressing 8 tokens to 1 cuts MFLOPs$^{'}$ by ${\sim}85\%$ (3.04 $\to$ 0.45) at only a 2.1\,pt Recall drop (0.624 $\to$ 0.603), a favorable trade when every retrieved candidate must be scored by the model.
Reinvesting the freed budget into a 128D-5L backbone with 1-token compression yields strictly better operating points: lower cost than 64D-3L at 8 tokens (1.91 vs.\ 3.04 MFLOPs$^{'}$) and higher Recall (0.657 vs.\ 0.624), confirming that compression-then-scale-up dominates more-tokens-with-smaller-backbone.

\subsubsection{Effect of Multi-Domain Modeling}
\label{sec:multidomain}

\begin{table}[t]
    \caption{Ablation on multi-domain modeling strategies (Industrial, R@2000 per domain). FLOPs are user-side only, since target-side processing is identical across variants.}
    \label{tab:multidomain}
    \setlength{\tabcolsep}{3pt}
    \centering
    \small
    \begin{tabular}{lcccccc}
    \toprule
    \textbf{Method} & \textbf{A} & \textbf{B} & \textbf{C} & \textbf{D} & \textbf{Overall} & \textbf{MFLOPs} \\
    \midrule
    \multicolumn{7}{l}{\textit{Per-domain training (domain-specific data only)}} \\
    Single Domain Data & 0.591 & 0.539 & 0.537 & 0.647 & 0.579 & 41.3 \\
    \midrule
    \multicolumn{7}{l}{\textit{Unified training (all domains' data merged)}} \\
    0 Domain Tokens & 0.254 & 0.276 & 0.251 & 0.289 & 0.265 & 41.3 \\
    1 Domain Token & 0.321 & 0.349 & 0.339 & 0.345 & 0.337 & 41.8 \\
    20 Domain Tokens & 0.431 & 0.451 & 0.460 & 0.473 & 0.452 & 49.8 \\
    TransRetrieval (ours) & \textbf{0.593} & \textbf{0.590} & \textbf{0.575} & \textbf{0.663} & \textbf{0.603} & 41.3 \\
    \bottomrule
    \end{tabular}
\end{table}

Effective multi-domain modeling necessitates capturing domain context under latency constraints.  Table \ref{tab:multidomain} evaluates different strategies using the 64D-3L configuration, reporting user-side FLOPs to explicitly quantify the overhead introduced by sequence expansion.
The results show that simply mixing multi-domain data without any domain signal (0 Domain Tokens) leads to severe performance degradation. Adding a single domain token (1 Domain Token) provides limited improvement but incurs extra computation. Repeating the domain token 20 times (20 Domain Tokens) further amplifies the domain signal and improves performance, but at a significantly higher computational cost (49.8 MFLOPs), which is prohibitive for large-scale deployment.
Our proposed method, injecting domain information as a position-style embedding, achieves the best overall performance (0.603 Recall@2000),
while maintaining the exact same inference cost as the 0 Domain Tokens baseline.
Additionally, the \textit{Single Domain Data} row further validates our multi-domain strategy: training on a single domain's data alone yields noticeably lower Recall than our unified model, confirming that cross-domain data sharing is a net positive.


\subsection{Online A/B Test Results (RQ4)}
\label{sec:online}

\begin{table}[t]
\centering
\setlength{\tabcolsep}{2pt}
\small
\caption{Online A/B Test Results of TransRetrieval.}
\label{tab:online_results}
\begin{tabular}{lccccccc}
\toprule
& \multicolumn{4}{c}{\textbf{Per-Domain Lift}} & \multicolumn{3}{c}{\textbf{Overall}} \\
\cmidrule(lr){2-5} \cmidrule(lr){6-8}
& A & B & C & D & Lift & 95\% Conf. Interval & p-value \\
\midrule
Revenue & 1.72\% & 2.51\% & 5.39\% & 0.87\% & 2.53\% & [2.22\%, 2.70\%] & $<0.0001$ \\
RPM & 1.22\% & 2.18\% & 3.69\% & 0.43\% & 1.28\% & [1.14\%, 1.56\%] & $<0.0001$ \\
\bottomrule
\end{tabular}
\end{table}

A month-long online A/B test was conducted within the Alibaba display advertising system on 5\% of production traffic, where the retrieval model is the only variable changed.
TransRetrieval-128D5L serves at 230 QPS with P99 latency under 40\,ms, identical to the production baseline, so the gains come at no end-to-end latency cost. The lower per-engine QPS (230 vs.\ 300) reflects the higher per-target computational cost (1.91 vs.\ 0.69 MFLOPs$^{'}$) and is absorbed by horizontal scaling.

As shown in Table~\ref{tab:online_results}, TransRetrieval achieved a significant $2.53\%$ increase in Platform Revenue and a $1.28\%$ lift in Revenue Per Mille over the highly optimized Production Baseline. The daily gains remain consistent with no systematic decay over the month. These financial gains were achieved while maintaining a healthy ecosystem: user experience metrics remained stable, and the advertisers' Return on Investment (ROI) stayed comparable ($+0.06\%$).

\section{Discussion and Conclusion}
\textbf{Lessons learned.}
\emph{(i)~Input conditioning before architecture redesign}: the scaling breakthrough came not from architectural novelty but from a parameter-free one-line fix (weighted average) that restored homogeneous token norms.
\emph{(ii)~Compression as reallocation, not sacrifice}: compressing 8 target tokens to 1 frees 85\% FLOPs, and reinvesting them into depth/width (64D-3L$\to$128D-5L) yields +3.3\,pt Recall at lower total cost.
\emph{(iii)~Domain unification as a scaling asset}: beyond avoiding per-domain cost multiplication, position-style domain embeddings let sparse domains benefit from cross-domain transfer rather than starving in isolation.

\textbf{Conclusion.}
Together, these three designs unlock scaling laws for the retrieval stage: on a 40-billion-interaction industrial dataset and on the public KuaiRand benchmark, TransRetrieval delivers log-linear scaling ($R^2\!=\!0.82$/$0.88$), state-of-the-art recall, and a 2.53\% platform revenue lift in month-long online A/B tests.

\begin{acks}

This work was supported by National Natural Science Foundation of China (No.62472428), Public Computing Cloud, Renmin University of China, the fund for building world-class universities (disciplines) of Renmin University of China, Alibaba Group through the Alibaba Innovative Research Program, and the Qiushi Academic Project of Renmin University of China (RUC25QSDL123).
We sincerely appreciate Cheng Chen, Fei Wang, Houyuan Xiang, Huimin Yi, Jianan Chen, Jiawen Liao, Junchen Dong, Mingjie Liu, Rui Wang, Wei He, Wenchao Wang, Xingyu Wen, Yan Jiang, Yan Zhang, Yunlong Xu, Zhengxiong Zhou, Zhengyu Liu, Zhensong Yan, and Zhenyuan Lai for implementing the key components of the data, training, and inference infrastructure.

\end{acks}

\clearpage

\section*{GenAI Usage Disclosure} During the preparation of this manuscript, the authors used generative AI tools for text polishing and language editing. The authors reviewed and edited all AI-generated suggestions and take full responsibility for the content of this publication.

\bibliographystyle{ACM-Reference-Format}
\bibliography{ref}

@article{DBLP:journals/corr/abs-2010-11929,
  author       = {Alexey Dosovitskiy and
                  Lucas Beyer and
                  Alexander Kolesnikov and
                  Dirk Weissenborn and
                  Xiaohua Zhai and
                  Thomas Unterthiner and
                  Mostafa Dehghani and
                  Matthias Minderer and
                  Georg Heigold and
                  Sylvain Gelly and
                  Jakob Uszkoreit and
                  Neil Houlsby},
  title        = {An Image is Worth 16x16 Words: Transformers for Image Recognition
                  at Scale},
  journal      = {CoRR},
  volume       = {abs/2010.11929},
  year         = {2020},
  url          = {https://arxiv.org/abs/2010.11929},
  eprinttype   = {arXiv},
  eprint       = {2010.11929},
  bibsource    = {dblp computer science bibliography, https://dblp.org}
}

@article{DBLP:journals/corr/abs-2203-15556,
  author       = {Jordan Hoffmann and
                  Sebastian Borgeaud and
                  Arthur Mensch and
                  Elena Buchatskaya and
                  Trevor Cai and
                  Eliza Rutherford and
                  Diego de Las Casas and
                  Lisa Anne Hendricks and
                  Johannes Welbl and
                  Aidan Clark and
                  Tom Hennigan and
                  Eric Noland and
                  Katie Millican and
                  George van den Driessche and
                  Bogdan Damoc and
                  Aurelia Guy and
                  Simon Osindero and
                  Karen Simonyan and
                  Erich Elsen and
                  Jack W. Rae and
                  Oriol Vinyals and
                  Laurent Sifre},
  title        = {Training Compute-Optimal Large Language Models},
  journal      = {CoRR},
  volume       = {abs/2203.15556},
  year         = {2022},
  url          = {https://doi.org/10.48550/arXiv.2203.15556},
  doi          = {10.48550/ARXIV.2203.15556},
  eprinttype    = {arXiv},
  eprint       = {2203.15556},
  bibsource    = {dblp computer science bibliography, https://dblp.org}
}

@inproceedings{DBLP:conf/icml/ZhangLCNLLZHYWP24,
  author       = {Buyun Zhang and
                  Liang Luo and
                  Yuxin Chen and
                  Jade Nie and
                  Xi Liu and
                  Shen Li and
                  Yanli Zhao and
                  Yuchen Hao and
                  Yantao Yao and
                  Ellie Dingqiao Wen and
                  Jongsoo Park and
                  Maxim Naumov and
                  Wenlin Chen},
  title        = {Wukong: Towards a Scaling Law for Large-Scale Recommendation},
  booktitle    = {Forty-first International Conference on Machine Learning, {ICML} 2024,
                  Vienna, Austria, July 21-27, 2024},
  publisher    = {OpenReview.net},
  year         = {2024},
  url          = {https://openreview.net/forum?id=8iUgr2nuwo},
  bibsource    = {dblp computer science bibliography, https://dblp.org}
}

@inproceedings{DBLP:conf/icml/ZhaiLLWLCGGGHLS24,
  author       = {Jiaqi Zhai and
                  Lucy Liao and
                  Xing Liu and
                  Yueming Wang and
                  Rui Li and
                  Xuan Cao and
                  Leon Gao and
                  Zhaojie Gong and
                  Fangda Gu and
                  Jiayuan He and
                  Yinghai Lu and
                  Yu Shi},
  title        = {Actions Speak Louder than Words: Trillion-Parameter Sequential Transducers
                  for Generative Recommendations},
  booktitle    = {Forty-first International Conference on Machine Learning, {ICML} 2024,
                  Vienna, Austria, July 21-27, 2024},
  publisher    = {OpenReview.net},
  year         = {2024},
  url          = {https://openreview.net/forum?id=xye7iNsgXn},
  bibsource    = {dblp computer science bibliography, https://dblp.org}
}

@inproceedings{DBLP:conf/cikm/XuWGGXHWL25,
  author       = {Songpei Xu and
                  Shijia Wang and
                  Da Guo and
                  Xianwen Guo and
                  Qiang Xiao and
                  Bin Huang and
                  Guanlin Wu and
                  Chuanjiang Luo},
  editor       = {Meeyoung Cha and
                  Chanyoung Park and
                  Noseong Park and
                  Carl Yang and
                  Senjuti Basu Roy and
                  Jessie Li and
                  Jaap Kamps and
                  Kijung Shin and
                  Bryan Hooi and
                  Lifang He},
  title        = {Climber: Toward Efficient Scaling Laws for Large Recommendation Models},
  booktitle    = {Proceedings of the 34th {ACM} International Conference on Information
                  and Knowledge Management, {CIKM} 2025, Seoul, Republic of Korea, November
                  10-14, 2025},
  pages        = {6193--6200},
  publisher    = {{ACM}},
  year         = {2025},
  url          = {https://doi.org/10.1145/3746252.3761561},
  doi          = {10.1145/3746252.3761561},
  bibsource    = {dblp computer science bibliography, https://dblp.org}
}

@inproceedings{DBLP:conf/recsys/CovingtonAS16,
  author       = {Paul Covington and
                  Jay Adams and
                  Emre Sargin},
  editor       = {Shilad Sen and
                  Werner Geyer and
                  Jill Freyne and
                  Pablo Castells},
  title        = {Deep Neural Networks for YouTube Recommendations},
  booktitle    = {Proceedings of the 10th {ACM} Conference on Recommender Systems, Boston,
                  MA, USA, September 15-19, 2016},
  pages        = {191--198},
  publisher    = {{ACM}},
  year         = {2016},
  url          = {https://doi.org/10.1145/2959100.2959190},
  doi          = {10.1145/2959100.2959190},
  bibsource    = {dblp computer science bibliography, https://dblp.org}
}

@inproceedings{DBLP:conf/cikm/ZhuFZJWHDWZGYCC25,
  author       = {Jie Zhu and
                  Zhifang Fan and
                  Xiaoxie Zhu and
                  Yuchen Jiang and
                  Hangyu Wang and
                  Xintian Han and
                  Haoran Ding and
                  Xinmin Wang and
                  Wenlin Zhao and
                  Zhen Gong and
                  Huizhi Yang and
                  Zheng Chai and
                  Zhe Chen and
                  Yuchao Zheng and
                  Qiwei Chen and
                  Feng Zhang and
                  Xun Zhou and
                  Peng Xu and
                  Xiao Yang and
                  Di Wu and
                  Zuotao Liu},
  editor       = {Meeyoung Cha and
                  Chanyoung Park and
                  Noseong Park and
                  Carl Yang and
                  Senjuti Basu Roy and
                  Jessie Li and
                  Jaap Kamps and
                  Kijung Shin and
                  Bryan Hooi and
                  Lifang He},
  title        = {RankMixer: Scaling Up Ranking Models in Industrial Recommenders},
  booktitle    = {Proceedings of the 34th {ACM} International Conference on Information
                  and Knowledge Management, {CIKM} 2025, Seoul, Republic of Korea, November
                  10-14, 2025},
  pages        = {6309--6316},
  publisher    = {{ACM}},
  year         = {2025},
  url          = {https://doi.org/10.1145/3746252.3761507},
  doi          = {10.1145/3746252.3761507},
  bibsource    = {dblp computer science bibliography, https://dblp.org}
}

@inproceedings{DBLP:conf/cikm/HanYCJJ0MHLJHZY25,
  author       = {Ruidong Han and
                  Bin Yin and
                  Shangyu Chen and
                  He Jiang and
                  Fei Jiang and
                  Xiang Li and
                  Chi Ma and
                  Mincong Huang and
                  Xiaoguang Li and
                  Chunzhen Jing and
                  Yueming Han and
                  MengLei Zhou and
                  Lei Yu and
                  Chuan Liu and
                  Wei Lin},
  editor       = {Meeyoung Cha and
                  Chanyoung Park and
                  Noseong Park and
                  Carl Yang and
                  Senjuti Basu Roy and
                  Jessie Li and
                  Jaap Kamps and
                  Kijung Shin and
                  Bryan Hooi and
                  Lifang He},
  title        = {{MTGR:} Industrial-Scale Generative Recommendation Framework in Meituan},
  booktitle    = {Proceedings of the 34th {ACM} International Conference on Information
                  and Knowledge Management, {CIKM} 2025, Seoul, Republic of Korea, November
                  10-14, 2025},
  pages        = {5731--5738},
  publisher    = {{ACM}},
  year         = {2025},
  url          = {https://doi.org/10.1145/3746252.3761565},
  doi          = {10.1145/3746252.3761565},
  bibsource    = {dblp computer science bibliography, https://dblp.org}
}

@inproceedings{DBLP:conf/cikm/BorisyukSZPAPBP24,
  author       = {Fedor Borisyuk and
                  Qingquan Song and
                  Mingzhou Zhou and
                  Ganesh Parameswaran and
                  Madhu Arun and
                  Siva Popuri and
                  Tugrul Bingol and
                  Zhuotao Pei and
                  Kuang{-}Hsuan Lee and
                  Lu Zheng and
                  Qizhan Shao and
                  Ali Naqvi and
                  Sen Zhou and
                  Aman Gupta},
  editor       = {Edoardo Serra and
                  Francesca Spezzano},
  title        = {LiNR: Model Based Neural Retrieval on GPUs at LinkedIn},
  booktitle    = {Proceedings of the 33rd {ACM} International Conference on Information
                  and Knowledge Management, {CIKM} 2024, Boise, ID, USA, October 21-25,
                  2024},
  pages        = {4366--4373},
  publisher    = {{ACM}},
  year         = {2024},
  url          = {https://doi.org/10.1145/3627673.3680091},
  doi          = {10.1145/3627673.3680091},
  bibsource    = {dblp computer science bibliography, https://dblp.org}
}

@article{DBLP:journals/corr/abs-2407-21022,
  author       = {Junjie Huang and
                  Jizheng Chen and
                  Jianghao Lin and
                  Jiarui Qin and
                  Ziming Feng and
                  Weinan Zhang and
                  Yong Yu},
  title        = {A Comprehensive Survey on Retrieval Methods in Recommender Systems},
  journal      = {CoRR},
  volume       = {abs/2407.21022},
  year         = {2024},
  url          = {https://doi.org/10.48550/arXiv.2407.21022},
  doi          = {10.48550/ARXIV.2407.21022},
  eprinttype    = {arXiv},
  eprint       = {2407.21022},
  bibsource    = {dblp computer science bibliography, https://dblp.org}
}

@article{DBLP:journals/pami/MalkovY20,
  author       = {Yury A. Malkov and
                  Dmitry A. Yashunin},
  title        = {Efficient and Robust Approximate Nearest Neighbor Search Using Hierarchical
                  Navigable Small World Graphs},
  journal      = {{IEEE} Trans. Pattern Anal. Mach. Intell.},
  volume       = {42},
  number       = {4},
  pages        = {824--836},
  year         = {2020},
  url          = {https://doi.org/10.1109/TPAMI.2018.2889473},
  doi          = {10.1109/TPAMI.2018.2889473},
  bibsource    = {dblp computer science bibliography, https://dblp.org}
}

@article{DBLP:journals/corr/BaKH16,
  author       = {Lei Jimmy Ba and
                  Jamie Ryan Kiros and
                  Geoffrey E. Hinton},
  title        = {Layer Normalization},
  journal      = {CoRR},
  volume       = {abs/1607.06450},
  year         = {2016},
  url          = {http://arxiv.org/abs/1607.06450},
  eprinttype    = {arXiv},
  eprint       = {1607.06450},
  bibsource    = {dblp computer science bibliography, https://dblp.org}
}

@inproceedings{DBLP:conf/kdd/MaZYCHC18,
  author       = {Jiaqi Ma and
                  Zhe Zhao and
                  Xinyang Yi and
                  Jilin Chen and
                  Lichan Hong and
                  Ed H. Chi},
  editor       = {Yike Guo and
                  Faisal Farooq},
  title        = {Modeling Task Relationships in Multi-task Learning with Multi-gate
                  Mixture-of-Experts},
  booktitle    = {Proceedings of the 24th {ACM} {SIGKDD} International Conference on
                  Knowledge Discovery {\&} Data Mining, {KDD} 2018, London, UK,
                  August 19-23, 2018},
  pages        = {1930--1939},
  publisher    = {{ACM}},
  year         = {2018},
  url          = {https://doi.org/10.1145/3219819.3220007},
  doi          = {10.1145/3219819.3220007},
  bibsource    = {dblp computer science bibliography, https://dblp.org}
}

@inproceedings{DBLP:conf/recsys/TangLZG20,
  author       = {Hongyan Tang and
                  Junning Liu and
                  Ming Zhao and
                  Xudong Gong},
  editor       = {Rodrygo L. T. Santos and
                  Leandro Balby Marinho and
                  Elizabeth M. Daly and
                  Li Chen and
                  Kim Falk and
                  Noam Koenigstein and
                  Edleno Silva de Moura},
  title        = {Progressive Layered Extraction {(PLE):} {A} Novel Multi-Task Learning
                  {(MTL)} Model for Personalized Recommendations},
  booktitle    = {RecSys 2020: Fourteenth {ACM} Conference on Recommender Systems, Virtual
                  Event, Brazil, September 22-26, 2020},
  pages        = {269--278},
  publisher    = {{ACM}},
  year         = {2020},
  url          = {https://doi.org/10.1145/3383313.3412236},
  doi          = {10.1145/3383313.3412236},
  bibsource    = {dblp computer science bibliography, https://dblp.org}
}

@inproceedings{DBLP:conf/cikm/ShengZZDDLYLZDZ21,
  author       = {Xiang{-}Rong Sheng and
                  Liqin Zhao and
                  Guorui Zhou and
                  Xinyao Ding and
                  Binding Dai and
                  Qiang Luo and
                  Siran Yang and
                  Jingshan Lv and
                  Chi Zhang and
                  Hongbo Deng and
                  Xiaoqiang Zhu},
  editor       = {Gianluca Demartini and
                  Guido Zuccon and
                  J. Shane Culpepper and
                  Zi Huang and
                  Hanghang Tong},
  title        = {One Model to Serve All: Star Topology Adaptive Recommender for Multi-Domain
                  {CTR} Prediction},
  booktitle    = {{CIKM} '21: The 30th {ACM} International Conference on Information
                  and Knowledge Management, Virtual Event, Queensland, Australia, November
                  1 - 5, 2021},
  pages        = {4104--4113},
  publisher    = {{ACM}},
  year         = {2021},
  url          = {https://doi.org/10.1145/3459637.3481941},
  doi          = {10.1145/3459637.3481941},
  bibsource    = {dblp computer science bibliography, https://dblp.org}
}

@inproceedings{DBLP:conf/nips/VaswaniSPUJGKP17,
  author       = {Ashish Vaswani and
                  Noam Shazeer and
                  Niki Parmar and
                  Jakob Uszkoreit and
                  Llion Jones and
                  Aidan N. Gomez and
                  Lukasz Kaiser and
                  Illia Polosukhin},
  editor       = {Isabelle Guyon and
                  Ulrike von Luxburg and
                  Samy Bengio and
                  Hanna M. Wallach and
                  Rob Fergus and
                  S. V. N. Vishwanathan and
                  Roman Garnett},
  title        = {Attention is All you Need},
  booktitle    = {Advances in Neural Information Processing Systems 30: Annual Conference
                  on Neural Information Processing Systems 2017, December 4-9, 2017,
                  Long Beach, CA, {USA}},
  pages        = {5998--6008},
  year         = {2017},
  url          = {https://proceedings.neurips.cc/paper/2017/hash/3f5ee243547dee91fbd053c1c4a845aa-Abstract.html},
  bibsource    = {dblp computer science bibliography, https://dblp.org}
}

@article{COPR,
  series     = {CIKM ’23},
  title      = {COPR: Consistency-Oriented Pre-Ranking for Online Advertising},
  url        = {http://dx.doi.org/10.1145/3583780.3615465},
  doi        = {10.1145/3583780.3615465},
  journal  = {Proceedings of the 32nd ACM International Conference on Information and Knowledge Management},
  publisher  = {ACM},
  address    = {Birmingham United Kingdom},
  author     = {Zhao, Zhishan and Gao, Jingyue and Zhang, Yu and Han, Shuguang and Lou, Siyuan and Sheng, Xiang-Rong and Wang, Zhe and Zhu, Han and Jiang, Yuning and Xu, Jian and Zheng, Bo},
  year       = {2023},
  month      = oct,
  pages      = {4974–4980},
  collection = {CIKM ’23}
}

@inproceedings{BAR,
  title     = {Bidding-Aware Retrieval for Multi-Stage Consistency in Online Advertising},
  author    = {Bin Liu and Yunfei Liu and Ziru Xu and Zhi Kou and Yeqiu Yang and Han Zhu and Jian Xu},
  booktitle = {Proceedings of the 35th ACM International Conference on Information and Knowledge Management (CIKM '26)},
  year      = {2026},
  publisher = {ACM},
  address   = {New York, NY, USA},
  doi       = {10.1145/3799682.3840091},
  url       = {https://doi.org/10.1145/3799682.3840091}
}

@article{DBLP:journals/corr/abs-2411-10057,
  author       = {Chi Liu and
                  Jiangxia Cao and
                  Rui Huang and
                  Kai Zheng and
                  Qiang Luo and
                  Kun Gai and
                  Guorui Zhou},
  title        = {KuaiFormer: Transformer-Based Retrieval at Kuaishou},
  journal      = {CoRR},
  volume       = {abs/2411.10057},
  year         = {2024},
  url          = {https://doi.org/10.48550/arXiv.2411.10057},
  doi          = {10.48550/ARXIV.2411.10057},
  eprinttype    = {arXiv},
  eprint       = {2411.10057},
  bibsource    = {dblp computer science bibliography, https://dblp.org}
}

@inproceedings{DBLP:conf/cikm/GaoLZCLL0022,
  author       = {Chongming Gao and
                  Shijun Li and
                  Yuan Zhang and
                  Jiawei Chen and
                  Biao Li and
                  Wenqiang Lei and
                  Peng Jiang and
                  Xiangnan He},
  editor       = {Mohammad Al Hasan and
                  Li Xiong},
  title        = {KuaiRand: An Unbiased Sequential Recommendation Dataset with Randomly
                  Exposed Videos},
  booktitle    = {Proceedings of the 31st {ACM} International Conference on Information
                  {\&} Knowledge Management, Atlanta, GA, USA, October 17-21, 2022},
  pages        = {3953--3957},
  publisher    = {{ACM}},
  year         = {2022},
  url          = {https://doi.org/10.1145/3511808.3557624},
  doi          = {10.1145/3511808.3557624},
  bibsource    = {dblp computer science bibliography, https://dblp.org}
}

@inproceedings{DBLP:conf/www/WangXHSZW25,
  author       = {Yihan Wang and
                  Fei Xiong and
                  Zhexin Han and
                  Qi Song and
                  Kaiqiao Zhan and
                  Ben Wang},
  editor       = {Guodong Long and
                  Michale Blumestein and
                  Yi Chang and
                  Liane Lewin{-}Eytan and
                  Zi Helen Huang and
                  Elad Yom{-}Tov},
  title        = {Unleashing the Potential of Two-Tower Models: Diffusion-Based Cross-Interaction
                  for Large-Scale Matching},
  booktitle    = {Proceedings of the {ACM} on Web Conference 2025, {WWW} 2025, Sydney,
                  NSW, Australia, 28 April 2025- 2 May 2025},
  pages        = {304--312},
  publisher    = {{ACM}},
  year         = {2025},
  url          = {https://doi.org/10.1145/3696410.3714829},
  doi          = {10.1145/3696410.3714829},
  bibsource    = {dblp computer science bibliography, https://dblp.org}
}

@inproceedings{DBLP:conf/www/ZhangDDLJG23,
  author       = {Yuan Zhang and
                  Xue Dong and
                  Weijie Ding and
                  Biao Li and
                  Peng Jiang and
                  Kun Gai},
  editor       = {Ying Ding and
                  Jie Tang and
                  Juan F. Sequeda and
                  Lora Aroyo and
                  Carlos Castillo and
                  Geert{-}Jan Houben},
  title        = {Divide and Conquer: Towards Better Embedding-based Retrieval for Recommender
                  Systems from a Multi-task Perspective},
  booktitle    = {Companion Proceedings of the {ACM} Web Conference 2023, {WWW} 2023,
                  Austin, TX, USA, 30 April 2023 - 4 May 2023},
  pages        = {366--370},
  publisher    = {{ACM}},
  year         = {2023},
  url          = {https://doi.org/10.1145/3543873.3584629},
  doi          = {10.1145/3543873.3584629},
  bibsource    = {dblp computer science bibliography, https://dblp.org}
}

@inproceedings{DBLP:conf/cikm/JiangLZYLXDZ22,
  author       = {Yuchen Jiang and
                  Qi Li and
                  Han Zhu and
                  Jinbei Yu and
                  Jin Li and
                  Ziru Xu and
                  Huihui Dong and
                  Bo Zheng},
  editor       = {Mohammad Al Hasan and
                  Li Xiong},
  title        = {Adaptive Domain Interest Network for Multi-domain Recommendation},
  booktitle    = {Proceedings of the 31st {ACM} International Conference on Information
                  {\&} Knowledge Management, Atlanta, GA, USA, October 17-21, 2022},
  pages        = {3212--3221},
  publisher    = {{ACM}},
  year         = {2022},
  url          = {https://doi.org/10.1145/3511808.3557137},
  doi          = {10.1145/3511808.3557137},
  bibsource    = {dblp computer science bibliography, https://dblp.org}
}

@inproceedings{DBLP:conf/cikm/ChenLZWLMHJXDZ22,
  author       = {Rihan Chen and
                  Bin Liu and
                  Han Zhu and
                  Yaoxuan Wang and
                  Qi Li and
                  Buting Ma and
                  Qingbo Hua and
                  Jun Jiang and
                  Yunlong Xu and
                  Hongbo Deng and
                  Bo Zheng},
  editor       = {Mohammad Al Hasan and
                  Li Xiong},
  title        = {Approximate Nearest Neighbor Search under Neural Similarity Metric
                  for Large-Scale Recommendation},
  booktitle    = {Proceedings of the 31st {ACM} International Conference on Information
                  {\&} Knowledge Management, Atlanta, GA, USA, October 17-21, 2022},
  pages        = {3013--3022},
  publisher    = {{ACM}},
  year         = {2022},
  url          = {https://doi.org/10.1145/3511808.3557098},
  doi          = {10.1145/3511808.3557098},
  bibsource    = {dblp computer science bibliography, https://dblp.org}
}

@inproceedings{DBLP:conf/cikm/LvCGZQZ0Z25,
  author       = {Xiao Lv and
                  Jiangxia Cao and
                  Shijie Guan and
                  Xiaoyou Zhou and
                  Zhiguang Qi and
                  Yaqiang Zang and
                  Ben Wang and
                  Guorui Zhou},
  editor       = {Meeyoung Cha and
                  Chanyoung Park and
                  Noseong Park and
                  Carl Yang and
                  Senjuti Basu Roy and
                  Jessie Li and
                  Jaap Kamps and
                  Kijung Shin and
                  Bryan Hooi and
                  Lifang He},
  title        = {{MARM:} Unlocking the Recommendation Cache Scaling-Law through Memory
                  Augmentation and Scalable Complexity},
  booktitle    = {Proceedings of the 34th {ACM} International Conference on Information
                  and Knowledge Management, {CIKM} 2025, Seoul, Republic of Korea, November
                  10-14, 2025},
  pages        = {2022--2031},
  publisher    = {{ACM}},
  year         = {2025},
  url          = {https://doi.org/10.1145/3746252.3761015},
  doi          = {10.1145/3746252.3761015},
  bibsource    = {dblp computer science bibliography, https://dblp.org}
}

@inproceedings{DBLP:conf/icdm/KangM18,
  author       = {Wang{-}Cheng Kang and
                  Julian J. McAuley},
  title        = {Self-Attentive Sequential Recommendation},
  booktitle    = {{IEEE} International Conference on Data Mining, {ICDM} 2018, Singapore,
                  November 17-20, 2018},
  pages        = {197--206},
  publisher    = {{IEEE} Computer Society},
  year         = {2018},
  url          = {https://doi.org/10.1109/ICDM.2018.00035},
  doi          = {10.1109/ICDM.2018.00035},
  bibsource    = {dblp computer science bibliography, https://dblp.org}
}

@inproceedings{DBLP:conf/cikm/SunLWPLOJ19,
  author       = {Fei Sun and
                  Jun Liu and
                  Jian Wu and
                  Changhua Pei and
                  Xiao Lin and
                  Wenwu Ou and
                  Peng Jiang},
  editor       = {Wenwu Zhu and
                  Dacheng Tao and
                  Xueqi Cheng and
                  Peng Cui and
                  Elke A. Rundensteiner and
                  David Carmel and
                  Qi He and
                  Jeffrey Xu Yu},
  title        = {BERT4Rec: Sequential Recommendation with Bidirectional Encoder Representations
                  from Transformer},
  booktitle    = {Proceedings of the 28th {ACM} International Conference on Information
                  and Knowledge Management, {CIKM} 2019, Beijing, China, November 3-7,
                  2019},
  pages        = {1441--1450},
  publisher    = {{ACM}},
  year         = {2019},
  url          = {https://doi.org/10.1145/3357384.3357895},
  doi          = {10.1145/3357384.3357895},
  bibsource    = {dblp computer science bibliography, https://dblp.org}
}

@article{recis,
  title        = {RecIS: Sparse to Dense, A Unified Training Framework for Recommendation Models},
  author       = {Zong, Hua and Zeng, Qingtao and Zhou, Zhengxiong and Han, Zhihua and Yan, Zhensong and Liu, Mingjie and Sun, Hechen and Liu, Jiawei and Hu, Yiwen and Wang, Qi and others},
  journal      = {arXiv preprint arXiv:2509.20883},
  year         = {2025}
}

\end{document}